\documentclass{nws}
\usepackage{amsopn}
\usepackage{subfig}
\usepackage{calc}
\usepackage{url}
\usepackage[utf8]{inputenc}

\usepackage{xcolor}
\definecolor{carolinablue}{HTML}{6699CC}

\newcommand{\sref}{\protect\subref}
\newcommand{\method}[1]{\texttt{\textsc{#1}}}

\DeclareMathOperator{\PPR}{PPR}

\DeclareMathOperator{\NCP}{NCP}
\DeclareMathOperator{\localNCP}{localNCP}
\DeclareMathOperator*{\argmin}{argmin}

\title{A local perspective on community structure in multilayer networks}

\author[L. G. S. Jeub et. al.]{LUCAS G. S. JEUB\\Oxford Centre for Industrial 
and Applied Mathematics, Mathematical Institute, University of Oxford, OX2 
6GG, UK \\
 MICHAEL W. MAHONEY,\\International Computer Science Institute, Berkeley, 
CA 94704\\Department of Statistics, University of California at Berkeley, 
Berkeley, CA 94720\\ PETER J. MUCHA\\Carolina Center for Interdisciplinary Applied Mathematics, Department of Mathematics, 
University of North Carolina, Chapel Hill, NC 27599-3250, USA\\MASON A. 
PORTER\\
Oxford Centre for Industrial and Applied Mathematics, Mathematical Institute, 
University of Oxford, OX2 6GG, UK\\
CABDyN Complexity Centre, University of Oxford, Oxford, OX1 1HP, UK\\
}

\begin{document}

\maketitle

\begin{abstract}
The analysis of multilayer networks is among the most active areas of network science, and there are now several methods to detect dense ``communities'' of nodes in multilayer networks. One way to define a community is as a set of nodes that trap a diffusion-like dynamical process (usually a random walk) for a long time. In this view, communities are sets of nodes that create bottlenecks to the spreading of a dynamical process on a network. We analyze the local behavior of different random walks on multiplex networks (which are multilayer networks in which different layers correspond to different types of edges) and show that they have very different bottlenecks that hence correspond to rather different notions of what it means for a set of nodes to be a good community. This has direct implications for the behavior of community-detection methods that are based on these random walks.

\end{abstract}

A ``community'' in a network describes a densely-connected set of nodes (often relative to a null model), and communities can reveal regularities in processes of network formation, strongly influence the behavior of dynamical processes that take place on a network, and be related to functional groups of nodes \cite{Porter2009,Fortunato2010,Coscia:2011jq}. One can examine community structure in a network from either a global perspective or a local one. When using a global perspective, one typically partitions a network into a set of (potentially overlapping) communities; by contrast, when taking a local perspective, one seeks to determine the community or communities associated with a given node (or set of nodes). A local perspective naturally allows the detection of overlapping communities, as local communities for different seed nodes can share nodes without having to be identical. 

From either a global or local perspective, communities can be viewed as dependent not only on network structure but also on dynamical processes (as a surrogate for function) on a network. Moreover, the choices of both dynamical process and initial conditions are very important \cite{Jeub:2015fc}.  A popular and successful approach for identifying community structure---both globally \cite{Rosvall:2008fi,Delvenne:2010vx} and locally \cite{Andersen:2006we,Leskovec:2009fy,Jeub:2015fc}---is to analyze the behavior of a diffusion, random walk, or other spreading process on a network. This exploits the connection between the presence of communities in a network and the behavior of associated dynamical processes on that network \cite{Lambiotte:2008,Lambiotte2009}. 

Although numerous tools have been developed for the analysis of networks \cite{Newman2010}, most of them concentrate on time-independent networks with only a single type of tie between entities. Such ordinary networks are often unable to capture the complex interactions among entities in the real world. In general, interactions (and the entities themselves) can change over time, and there can also be multiple types of interactions between the same pair of entities. Temporal networks allow one to examine the former situation \cite{Holme2012,holme2015}, and multiplex networks allow consideration of the latter \cite{faust}. The use of \emph{multilayer networks} \cite{Boccaletti:2014bz,Kivela:2014dm} allows one to examine either temporal networks or multiplex networks. In the former case, each layer represents a time or a time window (though it is important to think about issues such as discrete versus continuous time). In the latter case, each layer represents a type of interaction. One can also represent a multiplex temporal network by using a multilayer framework.

Because multilayer networks are graphical structures with nodes and edges, the notion of bottlenecks to dynamical processes on networks extends in a natural way to multilayer networks. (See Boccaletti \emph{et al.} \shortcite{Boccaletti:2014bz}, Kivel\"a \emph{et al.} \shortcite{Kivela:2014dm}, and Salehi \emph{et al.} \shortcite{salehi2014} for discussions of numerous dynamical processes on such networks.) The notion that diffusion-like dynamics should exhibit bottlenecks when there are good communities has been a fruitful perspective for generalizing algorithmic detection of global community structure from single-layer networks to multilayer networks \cite{Mucha:2010p2164,DeDomenico:2014vv}. In the present paper, we view community structure in multilayer networks from a local perspective, and we demonstrate that one can directly apply the methodology from Jeub \emph{et al.} \shortcite{Jeub:2015fc} by considering a dynamical process that traverses both intralayer and interlayer edges. (One can also study community structure in multilayer networks using other approaches, such as ones based on stochastic block models \cite{tiago2015}.) In particular, one way to do this is to define an appropriate random walk on a multilayer network.

In the present article, we illustrate some features that one can encounter as a consequence of the particular structure of multiplex networks. As examples, we use two different random walks to explore the structure of synthetic benchmark multiplex networks and two empirical multiplex networks. In Section~\ref{sec:random-walks}, we discuss random walks on multilayer networks in general and the two random walks that we choose to explore in more detail. In Section~\ref{sec:NCP}, we introduce our methodology for identifying and summarizing community structure in networks.  We then use this methodology to explore the behavior of the different random walks on synthetic benchmark networks in Section~\ref{sec:benchmark} and on a transportation and a social multiplex network in Section~\ref{sec:real-networks}. We conclude in Section \ref{discuss}.

%%%%

\section{Random Walks on Multilayer Networks}
\label{sec:random-walks}

As with a traditional network, different choices are possible when defining a random walk on a multilayer network \cite{Mucha:2010p2164,DeDomenico:2014uo,DeDomenico:2014vv,zhana2015}. Following \cite{Kivela:2014dm}, a \emph{multilayer network} $M(V_M, E_M, V, L)$ is a graph $G_M(V_M, E_M)$ with an additional layer structure. Here $V$ is a set of \emph{nodes}, $L$ is a set of \emph{layers}, $V_M\subseteq V \times L$ is a set of \emph{state nodes}\footnote{Following \cite{DeDomenico:2014vv}, we use the terms \emph{state node} to refer to a node-layer tuple and \emph{physical node} to refer to the collection of all state nodes that represent the same node.}, and $E_M\subseteq V_M \times V_M$ is a set of (directed) \emph{edges}. We use $i\alpha\in V_M$ to denote the state node that represents node $i\in V$ in layer $\alpha\in L$ and $(i\alpha, j\beta)\in E_M$ to denote a directed edge from state node $i\alpha$ to state node~$j\beta$. One can encode the connectivity structure of a multilayer network, including both intralayer and interlayer edges, using an \emph{adjacency tensor} $A$ (the analogue of the adjacency matrix for single-layer networks) with elements
\begin{equation}
	A^{i\alpha}_{j\beta} = \cases{1, &$(i\alpha, j\beta) \in E_M$\,, \cr  0\,,  & otherwise\,.}
\label{eq:adjacency}
\end{equation}

One can write a discrete-time random walk on a multilayer network as
\begin{equation}
	p_{i\alpha}(t+1)=\sum_{j\beta\in V_M} P_{i\alpha}^{j \beta} p_{j\beta}(t) \,, 
\label{eq:walk}
\end{equation}
where $p_{j\beta}(t)$ is the probability for a random walker to be at node $j$ in layer $\beta$ at time $t$ and $P_{i\alpha}^{j \beta}$ is the probability for a random walker at node $j$ in layer $\beta$ to transition to node $i$ in layer $\alpha$ in a time step. The \emph{transition tensor} $P$ encodes both the intralayer and interlayer behavior of the random walk. We also want the random walk to be ergodic, so that it has a well-defined stationary distribution $p_{i\alpha}(\infty)$. The stationary distribution is a fixed point of equation~\ref{eq:walk}. That is, it satisfies
\begin{equation}
	p_{i\alpha}(\infty) = \sum_{j\beta \in V_M} P_{i\alpha}^{j \beta} p_{j\beta}(\infty) \, .
\label{eq:stationary}
\end{equation}

There are different ways to define a random walk on a multilayer network that reduce to the usual definition of a random walk for a single-layer network. The most direct way to generalize the concept of a random walk to a multilayer network is the \emph{classical random walk} \cite{Mucha:2010p2164,DeDomenico:2014uo}, which treats interlayer edges and intralayer edges as equivalent objects (though they can be differentiated using heterogeneous spreading rates). The elements of the transition tensor for a classical random walk on a multilayer network are thus
\begin{equation}
	P^{i\alpha}_{j\beta} = \frac{A^{i\alpha}_{j\beta}}{\displaystyle\sum_{j \beta \in V_M} A^{i\alpha}_{j\beta}}\,.
\label{eq:classical_walk}
\end{equation}
An alternative way to generalize the concept of a random walk is by using a \emph{physical random walk}~\cite{DeDomenico:2014vv,DeDomenico:2014uo} with transition-tensor elements
\begin{equation}
	\tilde{P}^{i\alpha}_{j\beta} = \frac{A^{i\alpha}_{i\beta}}{\displaystyle\sum_{\beta\in L} A^{i\alpha}_{i\beta}} \frac{A^{i\beta}_{j\beta}}{\displaystyle\sum_{j\in V} A^{i\beta}_{j\beta}} \, .
\label{eq:physical_walk}
\end{equation}
One time step of this physical random walk corresponds to a random walker first switching layers with probabilities proportional to the weights of the interlayer edges and then performing an ordinary random-walk step in the new layer\footnote{Note that this definition of a physical random walk assumes that the multilayer network has diagonal coupling (i.e., that all interlayer edges are between state nodes that represent the same node).}. This type of physical random walk on a multilayer network is equivalent to a classical random walk on a transformed multilayer network with adjacency-tensor elements $\tilde{A}^{i\alpha}_{j\beta} = A^{i\alpha}_{i\beta} A^{i\beta}_{j\beta}$. This transformed multilayer network has non-diagonal, directed interlayer edges even when the original multilayer network is undirected.

\begin{figure}
\subfloat[Classical random walk]{\begin{minipage}{0.26\linewidth}\raisebox{-0.5\height}{\includegraphics[]{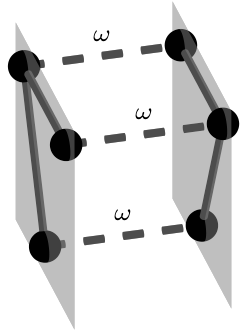}} \end{minipage}}\hfill
\subfloat[Relaxed random walk]{
\raisebox{-0.5\height}{\normalsize$(1-r)\;\times\;\!$}%
$\left(\raisebox{-0.5\height}{\includegraphics{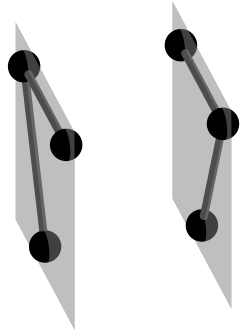}}\right)$%
\raisebox{-0.5\height}{\huge $\ + \  $}% 
\raisebox{-0.5\height}{\normalsize$r\; \times\;\!$}%
$\left(\raisebox{-0.5\height}{ \includegraphics[]{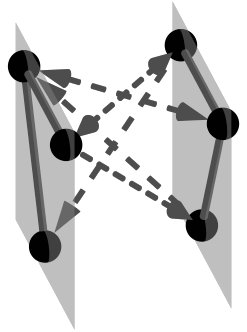}}\right)$}
\caption{\emph{Illustration of two types of random walk on a multilayer network.} The walks differ in the way that random walkers change layers. (a) \emph{Classical random walk} \protect\cite{Mucha:2010p2164,DeDomenico:2014uo}, in which we introduce interlayer edges with weight $\omega$ between state nodes (i.e., node-layer tuples) that represent the same physical node in adjacent layers. (b) \emph{Relaxed random walk} \protect\cite{DeDomenico:2014vv}, in which a random walker is constrained at each step to follow an edge within the same layer with probability $(1-r)$ and can choose any intralayer edge attached to the same physical node in any layer with probability $r$. In the latter case, the walker chooses uniformly at random from the set of all neighbors across all layers.
\label{fig:walks}}
\end{figure}

In this paper, we consider two types of random walks that have been proposed to study communities in multiplex networks: the classical random walk with uniform categorical coupling \cite{Mucha:2010p2164} and the relaxed random walk \cite{DeDomenico:2014vv}. For the classical random walk with uniform categorical coupling (which we henceforth call the ``classical random walk'' for short), we introduce interlayer edges with uniform weight $\omega \in [0,\infty]$ between all pairs of state nodes that correspond to the same physical node. That is, we define $A^{i\alpha}_{i\beta} = \omega$, $\alpha \neq \beta$ in Eq.~\ref{eq:classical_walk}. For the relaxed random walk, we constrain the random walker to follow an edge within the same layer with probability $1-r$ (so $r \in [0,1]$) and allow it to choose uniformly at random among all intralayer edges attached to the same physical node with probability $r$. Thus, the transition-tensor for the relaxed random walk has elements
\begin{equation}
	P^{i\alpha}_{j\beta} = (1-r) \delta(\alpha, \beta) \frac{A^{i\alpha}_{j\alpha}}{\displaystyle\sum_{j\in V} A^{i\alpha}_{j \alpha}} + r \frac{A^{i\beta}_{j\beta}}{\displaystyle \sum_{j \in V, \beta \in L} A^{i\beta}_{j\beta}} \, .
\end{equation}
Alternatively, one can think of the relaxed random walk as a physical random walk with the interlayer edges defined as 
\[
	A^{i\alpha}_{i\beta} = (1-r) \delta(\alpha, \beta) \sum_{j \in V, \alpha \in L} A_{j\alpha}^{i \alpha} + r \sum_{j \in V} A^{i \beta}_{j \beta} \, .
\]

In Fig.~\ref{fig:walks}, we illustrate the classical and relaxed random walks. In Section \ref{sec:NCP}, we discuss how we use random walks to identify communities and characterize mesoscale structures in multilayer networks.

%%%%%

\section{Local Communities and Network Community Profiles}
\label{sec:NCP}

We seek to contrast the behavior of different types of random walks on multilayer networks. From a dynamical perspective, communities correspond to sets of state nodes that create bottlenecks to a diffusive dynamical process. For a random walk, one measure to quantify bottlenecks is conductance\footnote{For a random walk on an undirected, single-layer network, this definition of conductance is equivalent to the conductance in \cite{Leskovec:2009fy,Jeub:2015fc}.} \cite{Jerrum:1988iv} 
\begin{equation}
	\phi(S)=\frac{\displaystyle \sum_{i\alpha\in S}\  \sum_{j\beta \notin S} P_{j\beta}^{i \alpha} p_{i\alpha}(\infty)}{\displaystyle\sum_{i\alpha\in S} p_{i\alpha}(\infty)}
	\label{eq:conductance}
\end{equation}
of a set $S\subset V_M$ of state nodes. Conductance measures the outflow of random walkers from a set of state nodes relative to the total number of random walkers within the set at stationarity. If a set of state nodes constitutes a bottleneck to a random walk, only a few of the random walkers present within the set should leave the set in a given time step, so the set should have low conductance. The two extreme cases are $\phi(S)=1$ if $S$ has no internal flow (i.e., no state node in $S$ is adjacent to any other state node in $S$) and $\phi(S)=0$ if $S$ is disconnected from the rest of a network. 

Different types of random walks correspond to different notions of what it means for a set of nodes to be a good community. Our choice of conductance as a measure of community quality is motivated by its nice theoretical properties. The presence of low-conductance sets (i.e., sets are considered to be ``good'' communities based on the conductance measure) relates directly to slow mixing of a random walk \cite{Mihail:1989vv}. There are efficient algorithms for identifying low-conductance sets with known approximation guarantees \cite{Andersen:2006we,Leskovec:2009fy}. There is also some empirical evidence that conductance is a reasonably effective measure for evaluating community quality and that other measures for evaluating community quality give reasonably similar results in practice \cite{Yang:2012tb}. However, conductance also has some limitations as a measure of community quality. Most notably, it is not very sensitive to the internal connectivity of putative communities. In the most extreme case, low-conductance sets may even be internally disconnected \cite{Leskovec:2009fy,Leskovec:2010uj,Jeub:2015fc}. Our choice of algorithm for identifying local communities (see our discussion below) somewhat mitigates this problem, as it implicitly optimizes the internal connectivity of the identified communities \cite{Leskovec:2010uj}.

We use the \method{ACLcut} method \cite{Andersen:2006we,Leskovec:2009fy,Jeub:2015fc} to identify putative communities. The \method{ACLcut} method is based on locally ranking nodes near a seed node by approximating a personalized \mbox{PageRank} ($\PPR$) score. Given an appropriate random walk (or other Markov process\footnote{More generally, it would also be both fruitful and interesting to develop local community-detection methods using dynamical processes that are not Markovian. A good start would be to use our approach through suitable adaptations of other processes that have been used to examine community structure in networks. Examples include Kuramoto phase oscillators \cite{arenas2006}; epidemic spreading processes \cite{Ghosh:2014wj}; and higher-order Markovian processes, such as those that have been employed in the study of ``memory networks'' \cite{renaud-natcomms}.}), one can define the associated $\PPR$ score of state node $i\alpha$ as the solution to the equation
\begin{equation}\label{ppr}
	\PPR(s,\gamma)_{i\alpha} = \gamma \sum_{j\beta\in V_M} P_{i\alpha}^{j \beta} 
\PPR(s,\gamma)_{j\beta} +(1-\gamma) s_{i\alpha}\,,
\end{equation}
where $s$ is a probability distribution that determines the seed nodes for the method \cite{Gleich:2014tn} and $\gamma \in [0,1]$ is a teleportation parameter. We use two different types of seeding procedure with the \method{ACLcut} method: 
\begin{itemize} 
\item seeding using a state node $j\beta$, where ${s_{i\alpha}=\delta(i\alpha,j\beta)}$ in Eq.~(\ref{ppr}); and \item seeding using a physical node $j$, where ${s_{i\alpha}=\delta(i,j)/|j|}$ in Eq.~(\ref{ppr}). \end{itemize}
 We describe the \method{ACLcut} method in more detail in Appendix~\ref{app:ACLcut}.

Our main tool for summarizing size-resolved community structure is a network community profile (NCP) \cite{Leskovec:2009fy}. An NCP shows the quality of the best community of a given size (i.e., number of nodes) as a function of community size. Because we are using conductance as a measure of community quality, we define the ``best'' community as the one with the lowest conductance. Hence, we define the NCP as 
\begin{equation}
\label{eq:NCP}
	\NCP(k) = \min_{S\subset V_M,\ |S| = k} \phi(S)\; .
\end{equation}
We also use local NCPs, where we constrain the communities to contain a given seed set $S_0$ of state nodes. That is,
\begin{equation}
\label{eq:localNCP}
	\localNCP(k,S_0) = \min_{S\subset V_M, \ |S| = k, \ S_0\subset S} \phi(S) \;.
\end{equation}
An NCP of a network can reveal interesting structural features about the network. In particular, its qualitative shape can reveal the global organization of a network \cite{Jeub:2015fc}. A local NCP is useful for identifying communities at different scales associated with a particular seed node (or seed set of nodes). 

Our code for identifying local communities and visualizing networks is available at \url{https://github.com/LJeub}. There is also a recently-proposed extension of the \method{ACLcut} method \cite{Kloster:2015ub} that allows one to sample local NCPs more efficiently, although we did not use it for our computations in this paper.

%%%%

\section{Synthetic Benchmark Multiplex Networks}
\label{sec:benchmark}

We now explore the behavior of the two different random walks on synthetic networks with known, planted community structure. The networks that we consider each have $n=1000$ nodes and $l=10$ layers (for a total of $10\;000$ state nodes, as we assume that every node exists on all layers) and $c=10$ planted communities. 

We sample the planted community structure in the different layers in the following way. We first sample a background community structure $S_b$ by sampling the community assignment for each node uniformly at random from $\{1,\ldots,c\}$. We then sample the community assignments $S_p$ for the state nodes such that a state node inherits the background community assignment of the corresponding physical node with probability $1-\lambda$ and otherwise its community assignment is sampled uniformly at random from $\{1,\ldots,c\}$. 

Given the community assignments for the state nodes, we sample the intralayer edges for the network independently from a block model, such that an edge between two state nodes in the same layer and with the same community assignment is present with probability $p_{in}$ and an edge between two state nodes in the same layer but with different community assignment is present with probability $p_{out}$. The ratio $p_{out}/p_{in}$ determines the strength of the community structure of the benchmark, where a small ratio indicates strong community structure. The parameter $\lambda$ controls the dependency between the layers; the layers have identical community structure for ${\lambda=0}$, and community structures in different layers are progressively less related to each other with increasing $\lambda$.

\begin{figure}
\begin{tabular}{p{0.09\linewidth}p{0.45\linewidth}p{0.01\linewidth}p{0.45\linewidth}}
&\multicolumn{1}{c}{Classical random walk} && \multicolumn{1}{c}{Relaxed random walk}\\
&\multicolumn{1}{c}{\includegraphics{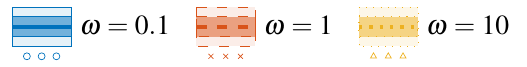}}&&\multicolumn{1}{c}{\includegraphics{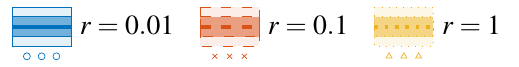}}\\
\hline
${\lambda=0}$&\raisebox{-\height+1\baselineskip}{\includegraphics[width=\linewidth]{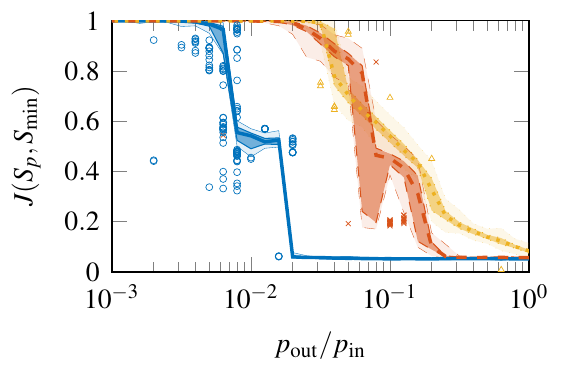}}&&
\raisebox{-\height+\baselineskip}{\includegraphics[width=\linewidth]{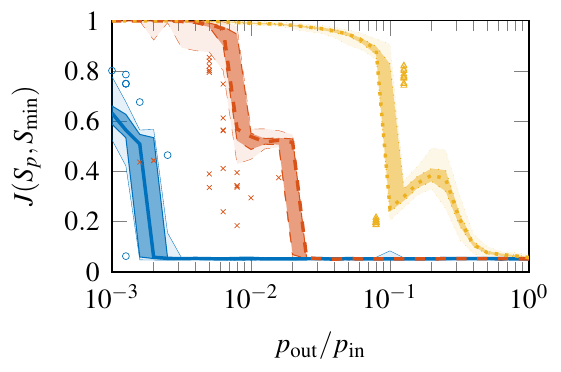}}\\
\hline
${\lambda=0.1}$&\raisebox{-\height+1\baselineskip}{\includegraphics[width=\linewidth]{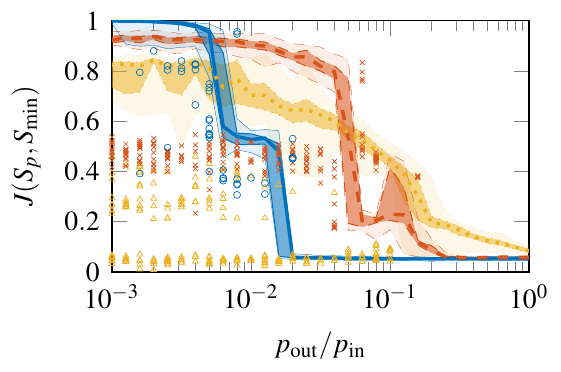}}&&
\raisebox{-\height+1\baselineskip}{\includegraphics[width=\linewidth]{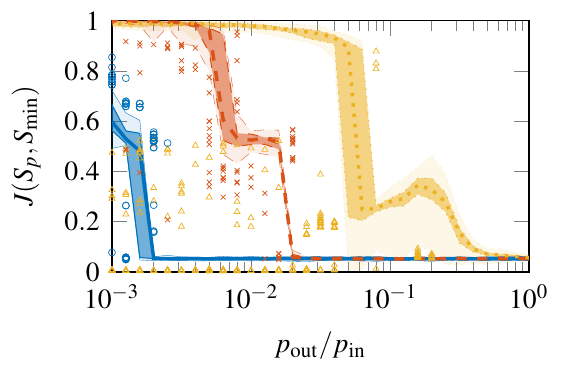}}\\
\hline
${\lambda=0.2}$&\raisebox{-\height+1\baselineskip}{\includegraphics[width=\linewidth]{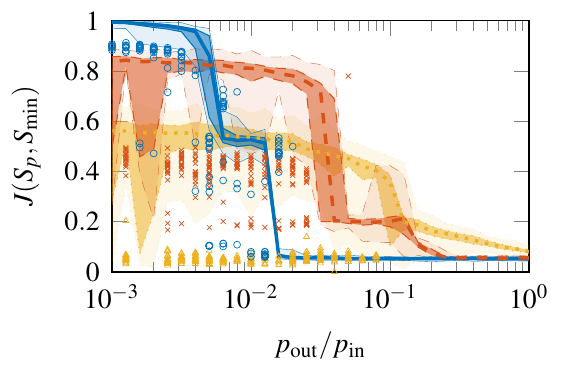}}&&
\raisebox{-\height+1\baselineskip}{\includegraphics[width=\linewidth]{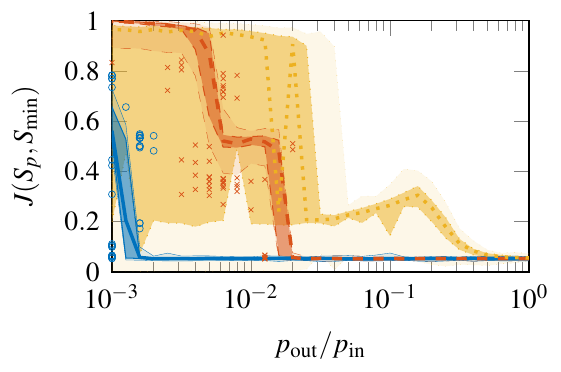}}\\
\hline
${\lambda=0.3}$&\raisebox{-\height+1\baselineskip}{\includegraphics[width=\linewidth]{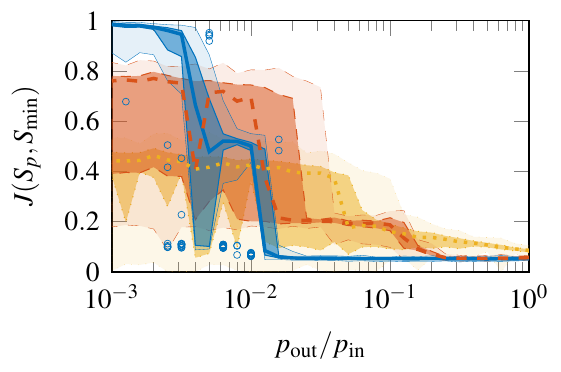}}&&
\raisebox{-\height+1\baselineskip}{\includegraphics[width=\linewidth]{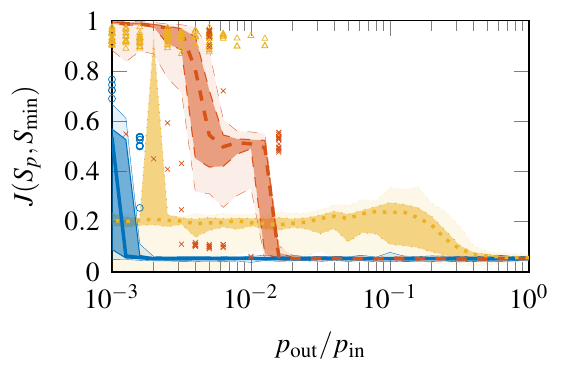}}
\end{tabular}

\caption{\emph{Recoverability of planted community structure using local NCPs.} For each value of $p_{\mathrm{out}}/p_{\mathrm{in}}$, we sample local NCPs for 100 uniformly-random seed nodes and compare the best community $S_{\min}$ identified by the local NCP to the planted community $S_p$ for the seed node using the Jaccard coefficient. The curves indicate the median of the Jaccard-coefficient distributions, the dark shaded regions indicate the second and third quartiles of the distributions, and the light shaded regions indicate the bulk of the distributions. Markers indicate outliers. \label{fig:bench-jaccard}} 
\end{figure}

In Fig.~\ref{fig:bench-jaccard}, we illustrate the ability of our local community-detection methods to recover the planted structure as we vary $p_{out}$ and $\lambda$. We fix ${p_{in}=0.1}$, ${n=1000}$, and ${l=c=10}$. We use local NCPs to identify communities in the following way. First, we select a state node $i\alpha$ uniformly at random as a seed node for the local NCP. We then identify the best community for $i\alpha$ as the community that achieves the minimum conductance 
\[
	S_{\min}(i\alpha) = \argmin_{S\subset V_M,\ i\alpha \in S} \phi(S) \;.
\]
among all communities that contain the state node $i\alpha$. This construction throws away a lot of information, as local minima of a local NCP can reveal interesting aspects of community structure. However, examining only globally optimal communities for a seed node enables one to easily compare algorithmically-obtained community structure with the planted structure. To compare the performance of the different random walks, we use the Jaccard coefficient \cite{Jaccard:1912bl}
 \[
	J\left(S_p(i\alpha), S_{\min}(i\alpha)\right)=\frac{\left|S_p(i\alpha) \cap S_{\min}(i\alpha)\right|}{\left|S_p(i\alpha) \cup S_{\min}(i \alpha)\right|}
\]
between the planted community and the best identified community for the seed node. (We obtain the same qualitative results when we compute normalized mutual information.) In Fig.~\ref{fig:bench-jaccard}, we show the distributions of the Jaccard coefficients for samples of 100 seed nodes. 

As we can see from Fig.~\ref{fig:bench-jaccard}, the performances of the two different random walks are comparable to each other; neither one is clearly better than the other. One interesting result is that, as we increase $\lambda$, it is increasingly pronounced that there are ``good'' and ``bad'' seed nodes for identifying community structure. For $\lambda=0$, the variability in the Jaccard coefficient for different seed nodes is fairly small, but as we increase $\lambda$, the number of outliers increases and we observe increasing variability in the distribution of the Jaccard coefficients. 

For small $\lambda$, strong interlayer coupling (i.e., for large values of $\omega$ or $r$, so that the rate of switching layers is high) helps identify the planted partition. For large enough values of $\lambda$, there is a range of $p_{\mathrm{out}}$ for which community structure is sufficiently strong that random walks with weaker interlayer coupling can outperform those with stronger interlayer coupling. This is already true when $\lambda=0.1$, and the difference becomes more pronounced as one increases $\lambda$.

For strong interlayer coupling, the two types of random walks (i.e., the classical random walk with $\omega=10$ and the relaxed walk with $r=1$) lose their ability to identify the planted structure as we increase $\lambda$ in rather different ways. For small values of $\lambda$ (in particular, for $\lambda=0.1$), the classical random walk identifies the background community structure rather than the planted community structure. As one increases $\lambda$ further, its performance at detecting both background and planted community structure decreases gradually for all seed nodes. 

In contrast, the performance of the relaxed walk deteriorates in a different way. It loses the ability to identify the planted structure for progressively more choices of seed nodes as we increase $\lambda$, but it still performs remarkably well for some seed nodes even when $\lambda=0.3$. 

%%%%%%

\section{Empirical Multiplex Networks}
\label{sec:real-networks}

We now illustrate our methodology on two empirical multiplex networks.\footnote{Note that for many multilayer networks (and, in particular, for the example networks that we examine in this paper), data on the weights of interlayer edges are not explicitly available \cite{Kivela:2014dm}.}
Our first example is the European Airline Network \cite{Cardillo:2013gn}, a multiplex transportation network with 37 layers, where each layer includes the flights for a single airline. Our second example is the Lazega Law Firm network \cite{Lazega:1999el,Lazega:2001wr,Snijders:2006cg}, a multiplex social network with three layers that represent advice, friendship, and co-work relationships between partners and associates of a corporate law firm. In Table~\ref{tab:data}, we highlight some key properties of these two networks.

\begin{table}
\caption{Example network data sets}
\label{tab:data}
\begin{tabular}{p{0.29\textwidth}@{\hspace{0.05\textwidth}}p{0.18\textwidth}@{\hspace{0.025\textwidth}}p{0.18\textwidth}@{\hspace{0.025\textwidth}}p{0.24\textwidth}}
\hline\hline
& nodes & edges & layers\\
\hline
European Airline Network \cite{Cardillo:2013gn}  & 450 airports & 3558 (undirected, unweighted) & 37 airlines\\
\\
Lazega Law Firm Network \cite{Lazega:1999el,Lazega:2001wr,Snijders:2006cg}& 71 employees& 2223 (directed, unweighted)& 3 (advice, friendship, co-work) \\
\hline\hline
\end{tabular}
\end{table}

\begin{figure}
\subfloat[European Airline Network\label{sfig:NCP_airline_aggregate}]{%
\includegraphics[width=0.47\linewidth]{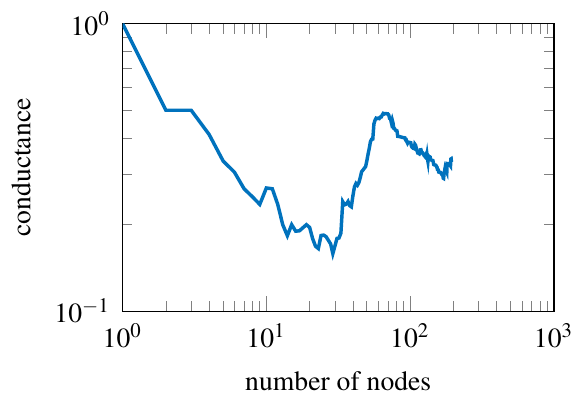}%
}%
\hfill
\subfloat[Lazega Law Firm Network\label{sfig:NCP_lazega_aggregate}]{%
\includegraphics[width=0.47\linewidth]{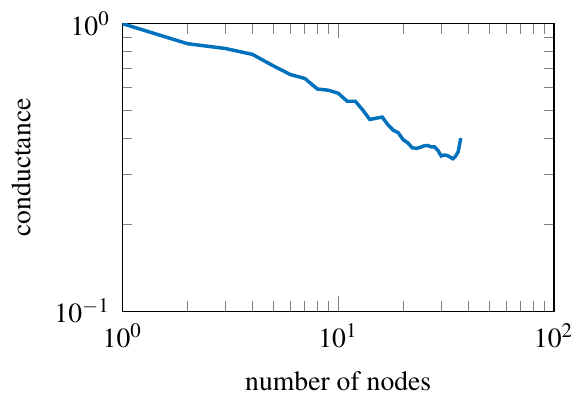}%
}

\caption{\emph{Network community profiles (NCPs) of two aggregated empirical networks.} We plot the quality (as measured by conductance) of the best community of each size (as measured by the number of nodes that are a member of the community).
\sref{sfig:NCP_airline_aggregate}~The NCP of the aggregate European Airline Network has a shape that one sees in networks with a core--periphery structure. \sref{sfig:NCP_lazega_aggregate}~The NCP of the aggregate Lazega Law Firm Network is slightly downward-sloping, and the high minimum conductance indicates that the aggregate network has no clear communities. \label{fig:NCPs_aggregate}}
\end{figure}

\begin{figure}
\subfloat[Classical random walk\label{sfig:NCP_airline_classical}]{%
\includegraphics[]{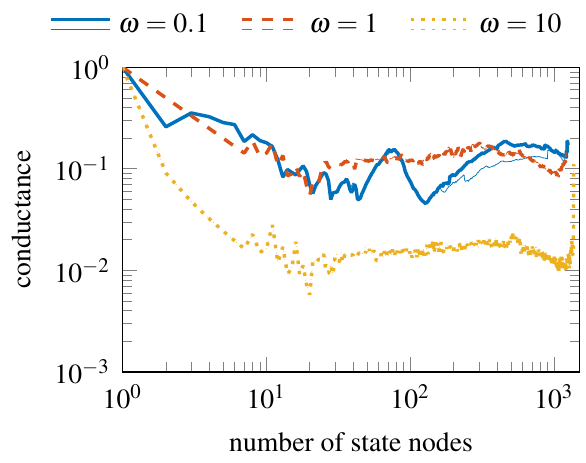}%
}%
\hfill
\subfloat[Relaxed random walk\label{sfig:NCP_airline_relaxed}]{%
\includegraphics[]{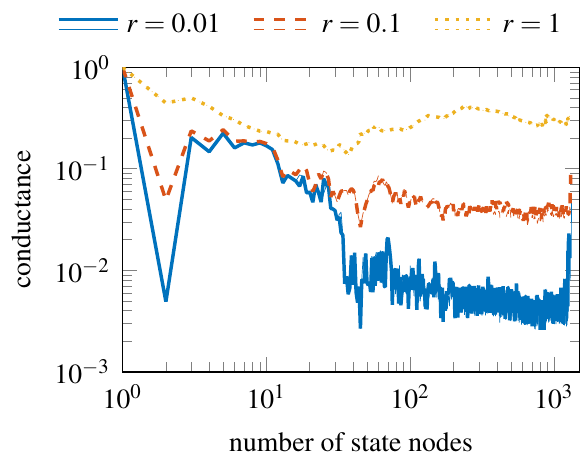}%
}

\subfloat[Best community with $173$ state nodes for ${\omega=0.1}$ (physical node as seed)]{%
\includegraphics[]{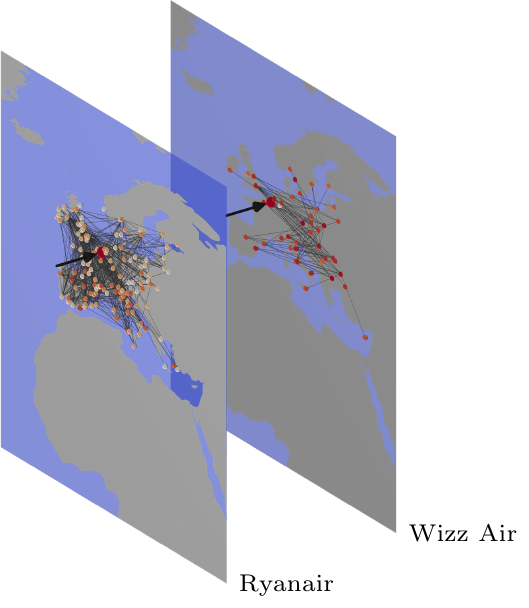}%
}%
\hfill
\subfloat[Best community with $169$ state nodes for ${r=0.1}$ (state node as seed)]{%
\includegraphics[]{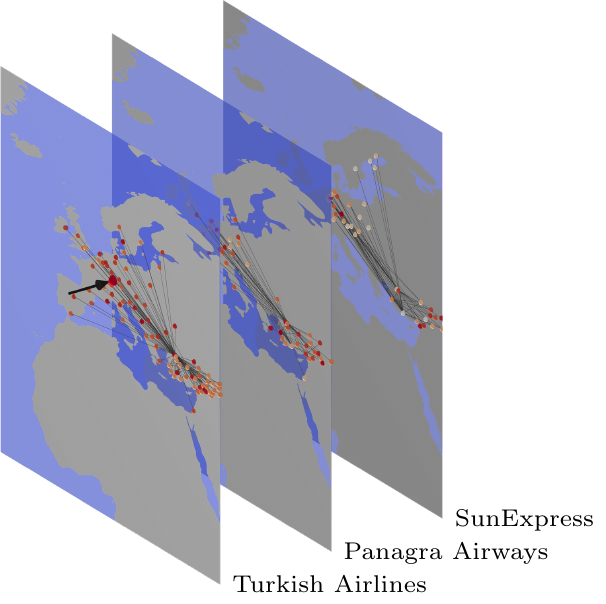}%
}

\caption{\emph{European Airline Network}. Panels (a) and (b) show NCPs for this network. We plot the quality (as measured by conductance) of the best community of each size (as measured by the number of state nodes that are a member of the community). Sampling using physical nodes (thin curves) versus using state nodes (thick curves) leads to very similar results, and the thin curves are typically hidden underneath the thick curves in this example. Panels (c) and (d) illustrate some of the communities that we obtain. We shade the state nodes in a community from dark red to light gray based on their rank (of their corresponding component) in the degree-normalized PPR-vector that we use to identify the community. (See Appendix~\ref{app:ACLcut} for details.) The large arrows point to the seed nodes. For small layer-jumping probability $r$ in the relaxed random walk and small interlayer edge weight $\omega$ in the classical random walk, the best communities tend to consist of sets of similar types of airlines (e.g., they fly to the same airport, are low-cost airlines, or share some other feature). The prominent dips in the NCPs in panel (b) for communities consisting of two state nodes are the result of a spurious connection in the data set that creates a bottleneck for the relaxed random walk. Even for ${r=1}$, the relaxed walk still predominantly identifies this type of community. By contrast, for large values of $\omega$, the classical random walk finds relatively geographically localized communities.
}
\label{fig:airline}
\end{figure}

\begin{figure}
\subfloat[Classical random walk\label{sfig:lz_NCP_c}]{%
\includegraphics[]{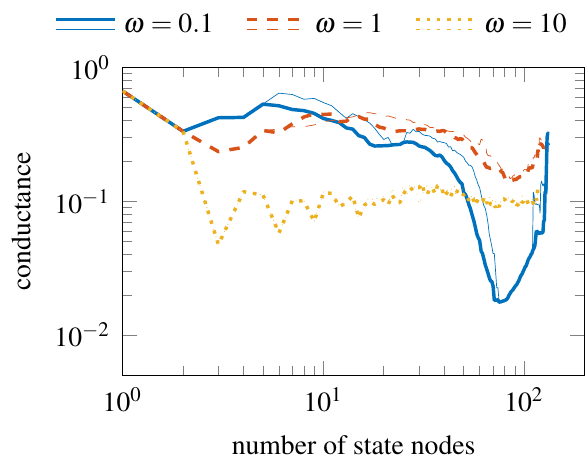}%
}%
\hfill
\subfloat[Relaxed random walk\label{sfig:lz_NCP_r}]{%
\includegraphics[]{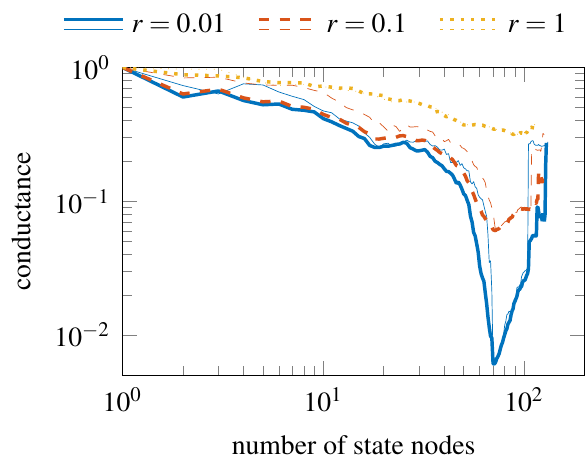}%
}

\subfloat[Best community with $19$ state nodes for ${\omega=0.1}$ (state node as seed)\label{sfig:lz_w0.1_n19}]{%
\includegraphics[]{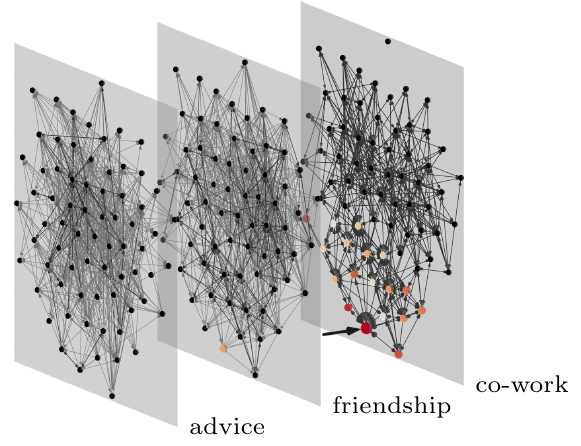}%
}%
\hfill
\subfloat[Best community with $19$ state nodes for ${r=0.01}$ (state node as seed)\label{sfig:lz_r0.01_n19}]{%
\includegraphics[]{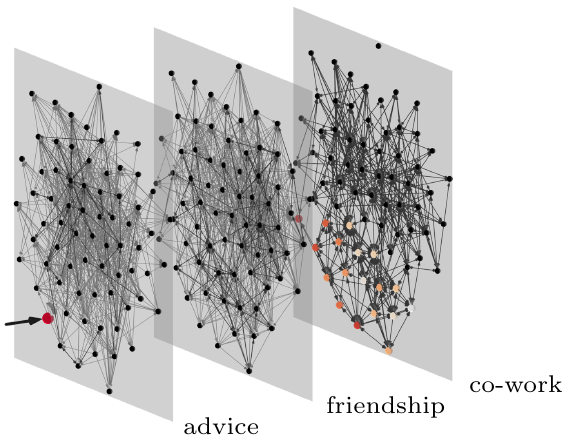}%
}

\caption{\emph{Lazega Law Firm Network}. Panels~\sref{sfig:lz_NCP_c}~and~\sref{sfig:lz_NCP_r} show NCPs for this network. As with the European Airline Network, when we use a small layer-jumping probability $r$ in the relaxed random walk and a small interlayer edge weight $\omega$ in the classical random walk, we obtain similar results even though we consider two different dynamical processes. We also again obtain similar results whether we use a state node or a physical node as a seed. For both types of random walks, with our choice of interlayer connection probability, the communities tend to be localized to a single layer. The prominent minimum in the NCPs at 71 nodes is the result of a community that contains all state nodes in the ``co-work'' layer. The communities that we highlight in panels~\sref{sfig:lz_w0.1_n19}~and~\sref{sfig:lz_r0.01_n19} are responsible for the other, less-pronounced minima in the NCPs at 19 nodes. They contain the members of the firm who are based in the Hartford office.
}
\label{fig:lazega}
\end{figure}

In Fig.~\ref{fig:NCPs_aggregate}, we show the NCPs of aggregated networks that we construct from our example multiplex networks. We define the weight of an edge between two nodes in an aggregated network as the number of edges in the corresponding multilayer network between the associated state nodes. That is, the adjacency matrix $\bar{A}$ of the aggregate network has entries
\[
	\bar{A}_{ij} = \sum_{\alpha \in L} A^{j\alpha}_{i\alpha} \;.
\]
The plots in Fig.~\ref{fig:NCPs_aggregate} give a reference for the NCPs of the multilayer networks (see Figs.~\ref{fig:airline} and~\ref{fig:lazega}). At the aggregate level, the European Airline Network has an NCP that is suggestive of a core--periphery structure (see \cite{csermely2013} for a review of such structure), although one cannot conclude this with certainty because the network is small and the conductance values are large. We do not observe any clear structure (and, in particular, no clear community structure) in the Lazega Law Firm Network.

\begin{sidewaysfigure}
\begin{minipage}[b][\textwidth][b]{\textheight}
\setcounter{subfigure}{\value{subfigure@save}}
\continuedfigure

\subfloat[Best community with $91$ state nodes for ${\omega=10}$ (physical node as seed)\label{sfig:lz_w10_n91}]{%
\includegraphics[]{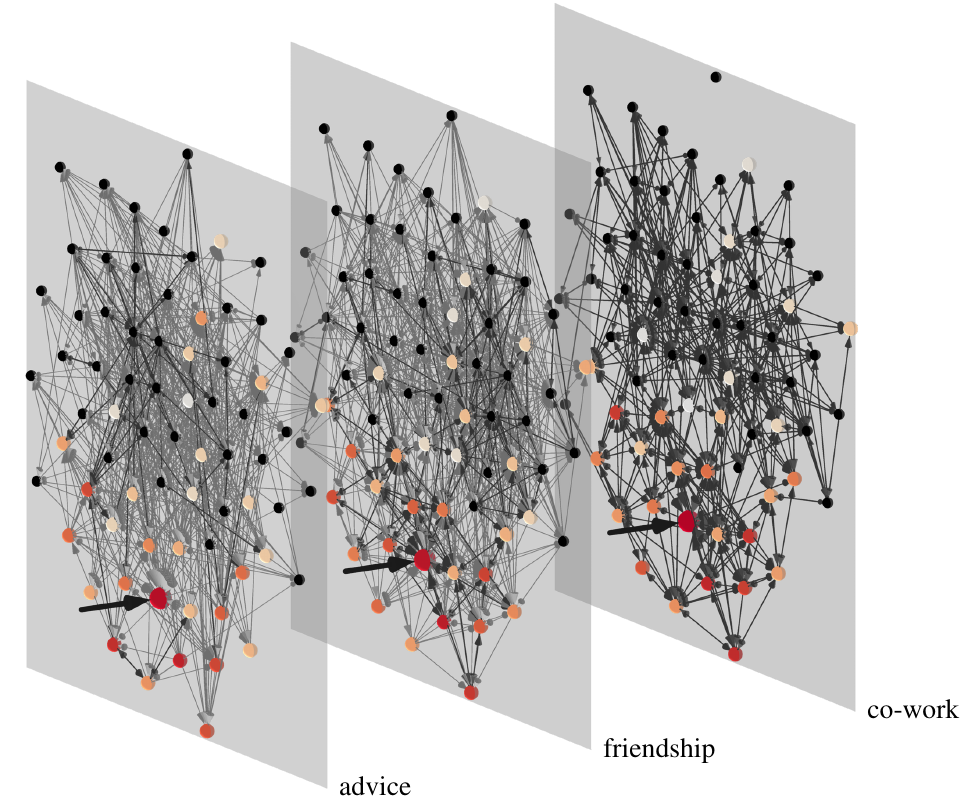}%
}%
\hfill
\subfloat[Best community with $91$ state nodes for ${r=1}$ (physical node as seed)\label{sfig:lz_r1_n91}]{%
\includegraphics[]{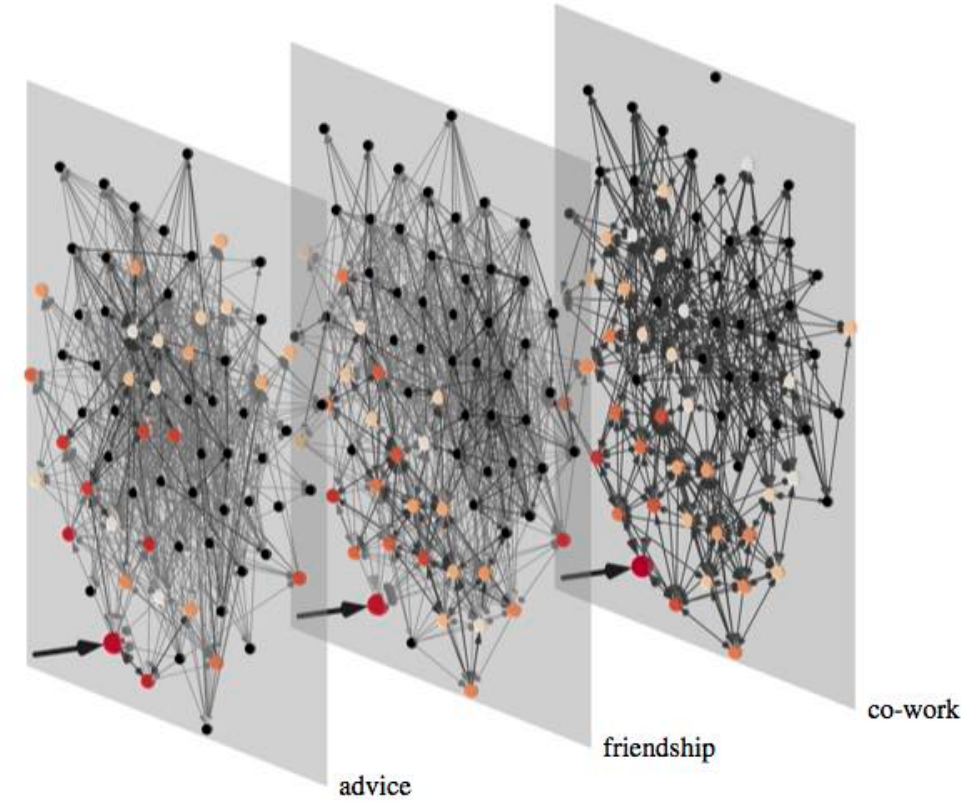}%
}

\caption{For large $\omega$, the classical random walk yields communities that are largely ``coherent'' across layers: if a state node is a member of a community, then the other state nodes associated with the same physical node also tend to be in that community. By contrast, communities from the relaxed random walk are not as coherent across layers.
 %compared to those from the classical walk.
 }
\end{minipage}
\end{sidewaysfigure}

In Figs.~\ref{fig:airline} and~\ref{fig:lazega}, we explore the multilayer structure of the airline and law-firm networks. An interesting aspect of multilayer networks is that one can use either physical nodes or state nodes as seeds to sample local communities. We compare the results of these two sampling procedures in Figs.~\ref{fig:airline} and~\ref{fig:lazega}.
%, where we use thin lines to indicate the results obtained by sampling using physical nodes. 
In these two networks, the two sampling procedures produce very similar results. In some cases, sampling using physical nodes can result in slightly better communities. (See the thick and thin solid blue curves in Fig.~\ref{sfig:NCP_airline_classical} for community sizes between $10^2$ to $10^3$ state nodes.) In other cases, sampling using physical nodes results in slightly worse communities. (See the thick and thin solid blue curves and dashed red curves in panels~\sref{sfig:lz_NCP_c} and~\sref{sfig:lz_NCP_r} of Fig.~\ref{fig:lazega}.)

For directed networks (e.g., the network in Fig.~\ref{fig:lazega}), random walks are not necessarily ergodic, so they might not have a unique stationary distribution. We use unrecorded edge teleportation \cite{Lambiotte:2011fe,DeDomenico:2014vv} to ensure that the random walk is ergodic. This corresponds to replacing the stationary distribution of the random walk in the definition of conductance (Eq.~\ref{eq:conductance}) by a PPR vector in which the seed vector is proportional to the vector of in-degrees of the state nodes. That is,
\[
	p_{i\alpha}(\infty)=\PPR(s,\gamma),\qquad s_{i\alpha}=\frac{\sum_{j\beta\in V_M} A^{j\beta}_{i\alpha}}{\sum_{i\alpha \in V_M} \sum_{j\beta\in V_M} A^{j\beta}_{i\alpha}} \;.
\]
For our results in Fig.~\ref{fig:lazega}, we use a teleportation rate of $\gamma=0.05$. For weighted networks in which intralayer weights are very different in different layers, one may also need to rescale the edge weights appropriately \cite{Cranmer:2014ut} to obtain results that are not dominated by a single layer (or small set of layers). However, this issue does not arise in our example networks in the present paper.

As one can see from Figs.~\ref{fig:airline} and~\ref{fig:lazega}, the NCPs for the multilayer networks look very different from those of the aggregated networks in Fig.~\ref{fig:NCPs_aggregate}. One exception are the NCPs for the relaxed walk with rate $r=1$; its shape resembles that of the NCPs for the aggregated network. For each of these networks, the multilayer structure introduces bottlenecks to the spreading of the random walks that are not present in the associated aggregated networks. This is also reflected in the types of communities that underlie these bottlenecks.

For the airline network, the best communities identified by random walks with weak interlayer coupling tend to contain all state nodes from a given layer or a set of layers. Effectively, the local communities are identifying sets of similar airlines that share many common destinations. However, as we illustrate in Fig.~\ref{fig:airline}, the exact communities identified by the classical random walk and relaxed random walk can be very different. Once the interlayer coupling becomes sufficiently strong, the nature of the communities identified by the classical random walk changes completely. The best communities identified by the classical random walk with strong interlayer coupling tend to be localized geographically and span all layers. The relaxed random walk, however, still predominately identifies sets of airlines even when $r=1$ (i.e., when the interlayer coupling is maximal). In fact, the communities identified by the relaxed random walk with $r=1$ are often rather similar to those identified with $r=0.1$. 

For the Lazega Law Firm Network, the layer structure also results in bottlenecks to the random walks when the interlayer coupling is weak. This is the cause for the sharp minima in the NCPs in Fig.~\ref{fig:lazega} for community sizes of 71 state nodes. At smaller community sizes, one can identify layer internal structures---most notably a community in the ``co-work'' layer that consists of the members of the firm that work in the Hartford office. Unlike for the airline network, in the case of the Lazega Law Firm Network, both types of random walk predominantly identify communities that span all layers when the interlayer coupling is sufficiently strong. However, the two types of random walks explore the law-firm network in rather different ways.  The classical random walk explores the different layers of the network in a ``coherent'' manner when the interlayer coupling is strong. That is, if a state node associated with a particular physical node is included in a community, then the other state nodes of that physical node (i.e., its manifestation in the other layers) tend to also belong to the community. However, as we illustrate in panels~\sref{sfig:lz_w10_n91} and~\sref{sfig:lz_r1_n91} of Fig.~\ref{fig:lazega}, the communities identified by the relaxed random walk tend to be less coherent across layers than those identified by the classical random walk.

To understand the difference in behavior between the relaxed and classical random walks at high layer-switching rates (i.e., for ${r\approx 1}$ and for large $\omega$), it is important examine the behavior of the two dynamical processes that are induced on the aggregated networks. The dynamical process induced on the aggregated network by the relaxed random walk with ${r=1}$ is simply a standard random walk. However, the process induced by the classical random walk for large $\omega$ explores the aggregated network much more slowly than a standard random walk, as most of the transitions are between state nodes that represent the same physical node. This results in a downward shift of the NCPs as one increases $\omega$. This has a similar effect as introducing a self-loop at each node. (As discussed in Arenas \emph{et al.} \shortcite{Arenas:2008hq}, introducing self-loops is one way to introduce a resolution parameter in the modularity quality function.) 

%%%%%%

\section{Discussion and Conclusions}\label{discuss}

We have seen using example synthetic networks that bottlenecks of random walks on a multilayer network can reveal nontrivial multiplex community structure in which community structure in different layers of a network is related but not identical. We explored two types of random walks --- a classical random walk and a relaxed random walk --- for identifying structure in our synthetic benchmarks, and we found that they can behave rather differently from each other in some situations. Consequently, different random walks give different community structures in a network, and one thus also expects to observe (although we did not test this directly) different community structures in different global methods (e.g., based on optimizing a quality function) based on the two different types of random walks. Similar results have been noted previously in other contexts \cite{Lambiotte:2008,Lambiotte2009}.

As we saw in Section~\ref{sec:real-networks}, the behavior of the random walks on a multilayer network is in general very different from the behavior of a random walk on a corresponding aggregated network. Consequently, examining a multilayer network can reveal important information that is not visible in a corresponding aggregated network. In particular, bottlenecks to random walks on a multilayer network can reveal structures, such as sets of related layers and communities confined to a single layer, that are impossible to see in aggregated networks. 

Our approach is very general, and a suite of other dynamical processes can also used to develop a diverse family of local community-detection methods. In addition to considering other processes, in advancing our work further, it will also be interesting to exploit transformations between ordinary random walks and other types of random walks \cite{lambiotte2011-flow,Yan:2016ky}. Another interesting extension of our approach would be to use it as the seed-set-expansion part of a seed-centric algorithm \cite{Kanawati:2014ha,Hmimida:2015hr,Whang:2016jz} for detecting communities in multilayer networks. 

%%%%%%

\section*{Acknowledgements}

LGSJ acknowledges a CASE studentship award from the EPSRC (BK/10/039), and LGSJ and MAP were supported in part from the James S. McDonnell Foundation 21st Century Science Initiative - Complex Systems Scholar Award grant \# 220020177 and the FET-Proactive project PLEXMATH (FP7-ICT-2011-8; grant \# 317614) funded by the European Commission. MAP was also supported by the EPSRC (EP/J001759/1). MWM acknowledges funding from the Army Research Office and from the Defense Advanced Research Projects Agency. PJM was supported from the James S. McDonnell Foundation 21st Century Science Initiative - Complex Systems Scholar Award grant \# 220020315. MAP also thanks SAMSI for supporting several visits and MWM for his hospitality during a sabbatical at Stanford. 

%%%

\appendix

\section{Sampling Network Community Profiles (the \method{ACLcut} method)}
\label{app:ACLcut}

\begin{figure}
\figrule
\programmath
\begin{tabular}{l@{\hspace{2em}}l}
$\mathbf{def}\ \method{ACLcut}(P,v,S_0, \gamma,\epsilon):$& \\[1ex]
$\qquad \mathbf{for}\ i\alpha \in V_M:$&set up seed vector\\[1ex]
$\qquad \qquad \mathbf{if}\ i\alpha \in S_0:$&\\
$\qquad \qquad \qquad s_{i\alpha} = 1/|S_0|$&\\[1ex]
$\qquad \qquad \mathbf{else} :$& \\
$\qquad \qquad \qquad s_{i\alpha} = 0$&\\[1ex]
$\qquad p=\method{APPR}(P,v,s,\gamma,\epsilon)$& compute $\epsilon$-approximate PageRank vector\\
$\qquad \phi, \mathcal{S} = \method{SweepCut}(P,v,p/v)$& normalized sweep cut\\
$\qquad \mathbf{return}\ \phi,\mathcal{S}$&return conductance and communities
\end{tabular}
\unprogrammath
\caption{\method{ACLcut} method for sampling local communities. The inputs are the transition tensor $P$ of the random walk, a seed set $S_0$ of state nodes, and a vector $v$ of node volumes --- which is proportional either to the stationary distribution of $P$ or (when considering teleportation) to a PageRank vector. The resolution of the method is controlled by the teleportation parameter $\gamma$ and the truncation parameter $\epsilon$. \label{alg:ACLcut}}
\figrule
\end{figure}

\begin{figure}
\figrule
\programmath
\begin{tabular}{l@{\hspace{2em}}l}
$\mathbf{def}\ \method{APPR}(P,v,s,\gamma,\epsilon) :$&\\
$\qquad \tilde{\gamma}=(1-\gamma)/(1+\gamma)$ & convert to equivalent lazy-walk teleportation\\[1ex]
$\qquad \mathbf{for} \ i\alpha \in V_M :$&\\
$\qquad \qquad p_{i\alpha}=0$& initialize PageRank vector\\
$\qquad \qquad e_{i\alpha}=s_{i\alpha}$& initialize residual\\[1ex]
$\qquad Q=\{i\alpha: e_{i\alpha}\geq\epsilon v_{i\alpha}\}$&keep track of nodes to update\\[1ex]
$\qquad \mathbf{while} \ Q\neq \emptyset :$&\\
$\qquad \qquad i\alpha = \mathbf{pop}(Q)$& select a node to update\\
$\qquad \qquad \tilde{e} = e_{i\alpha}$&\\
$\qquad \qquad p_{i\alpha}=p_{i\alpha} + \tilde{\gamma} \tilde{e}$& push probability mass to PageRank vector\\
$\qquad \qquad e_{i\alpha} = (1-\tilde{\gamma})\tilde{e}/2$&\\[1ex]
$\qquad \qquad \mathbf{for} \ j\beta \in \{j\beta: P^{i\alpha}_{j\beta}>0\} :$&\\
$\qquad \qquad \qquad e_{j\beta}=e_{j\beta} + (1- \tilde{\gamma})P^{i\alpha}_{j\beta} \tilde{e}/2$&\\[1ex]
$\qquad \qquad Q=\{i\alpha: e_{i\alpha} \geq \epsilon v_{i\alpha}\}$ & check $i\alpha$ and its neighbors to update $Q$\\[1ex]
$\qquad \mathbf{return}\ p$& return $\epsilon$-approximate PageRank vector\\
\end{tabular}
\unprogrammath

\caption{\method{APPR} procedure for computing $\epsilon$-approximate PageRank vectors using only local information. See Fig.~\ref{alg:ACLcut} for a description of the input arguments.\label{alg:APPR}}
\figrule
\end{figure}

\begin{figure}
\figrule
\programmath
\begin{tabular}{l@{\hspace{2em}}l}
$\mathbf{def}\ \method{SweepCut}(P,v,p):$&\\
$\qquad N=\method{RankOrder}(V_M,p)$ & return state nodes in descending order of $p$\\
$\qquad S=\emptyset$&\\
$\qquad c=0$&\\
$\qquad vol=0$&\\[1ex]
$\qquad \mathbf{for} \ k \in \{1,\ldots,|N|\}:$&\\
$\qquad \qquad i\alpha=N(k)$& get the next state node to consider\\
$\qquad \qquad c=c-\sum_{j\beta \in S} P^{j\beta}_{i\alpha} v_{j\beta} + \sum_{j\beta\notin S} P^{i\alpha}_{j\beta} v_{i\alpha}$&update conductance\\
$\qquad \qquad vol= vol + v_{i\alpha}$&\\
$\qquad \qquad S=S\cup i\alpha$&\\
$\qquad \qquad \phi_k = c/vol$&\\
$\qquad \qquad \mathcal{S}(k) = S$&\\[1ex]
$\qquad \mathbf{return}\ \phi, \mathcal{S}$&return conductance values and sweep sets\\
\end{tabular}
\unprogrammath

\caption{\method{SweepCut} procedure for identifying communities based on a ranking vector $p$ for the state nodes.\label{alg:SweepCut}}
\figrule
\end{figure}

We use the \method{ACLcut} method \cite{Andersen:2006we,Leskovec:2009fy,Jeub:2015fc} to sample local communities and network community profiles (NCPs). In this appendix, we briefly discuss how we apply this procedure to identify communities using a general random walk on a multilayer network. 

The key idea behind the \method{ACLcut} method is the use of a ``push'' procedure \cite{Andersen:2006we}, which pushes probability mass from the residual vector $e$ to the PageRank vector $p$ while preserving the invariant $p=\PPR(\gamma,s-e)$. We describe the different parts of the \method{ACLcut} method in Figs.~\ref{alg:ACLcut}--\ref{alg:SweepCut}. In addition to the teleportation parameter $\gamma$, the \method{ACLcut} method also depends on a truncation parameter $\epsilon$. The \method{ACLcut} method terminates once the residual is small enough so that $e_{i\alpha}<\epsilon v_{i\alpha}$ for all state nodes $i\alpha\in V_M$, where the quantity $v$ denotes a vector of node volumes. We set $v=|V_M|\times p(\infty)$, where $p(\infty)$ is either the stationary distribution of the random walk or (when considering teleportation) it is a PageRank vector. The rescaling is purely for computational convenience, as from a theoretical perspective the results are  invariant under rescaling of the node volumes (because one also rescales $\epsilon$). 

To sample an NCP, we use the \method{ACLcut} method to sample communities for different values of $\epsilon$ and different seed nodes, and we take the lower envelope of the conductance values. To sample a local NCP, we only vary $\epsilon$ and use the seed set of the local NCP as a seed set for the \method{ACLcut} method. We use 20 logarithmically spaced values for $\epsilon$ in the interval $[1/\max(v),1/\sum v]$, and we fix $\gamma=0.998$. For each value of $\epsilon$, we initially set the sample set $S$ of potential seed nodes to be either the set of all state nodes (i.e., $S=V_M$) or the set of all physical nodes (i.e., $S=V$). We then sample seed nodes uniformly at random without replacement from $S$ until $S$ is empty. To avoid excessive computations for small values of $\epsilon$, we remove nodes from $S$ once they have been included in the best local community returned by the \method{ACLcut} method 10 times. This sampling procedure allows one to estimate an NCP in a time that scales almost linearly with the number of state nodes while ensuring good coverage of the structure of a network.

%\bibliography{local_multiplex_r1_2}

\end{document}